\documentclass[a4paper,11pt]{article}
\usepackage{pos}

\title{First measurement of anti-k$_\mathrm{T}$ jet spectra and jet substructure using the archived ALEPH $e^+e^-$ data at 91.2 GeV}

\author*[a]{Yi Chen}
\author[b]{Austin Baty}
\author[c]{Dennis Perepelitsa}
\author[c]{Christopher McGinn}
\author[a]{Jesse Thaler}
\author[d]{Marcello Maggi}
\author[e]{Paoti Chang}
\author[a]{Tzu-An Sheng}
\author[f]{Yang-Ting Chien}
\author[a]{Yen-jie Lee}

\affiliation[a]{Massachusetts Institute of Technology\\
  77 Massachusetts Ave., Cambridge, MA 02139, USA}
\affiliation[b]{Rice University\\
  6100 Main St, Houston, TX 77005, USA}
\affiliation[c]{University of Colorado Boulder\\
  Boulder, CO 80309, USA}
\affiliation[d]{Istituto Nazionale di Fisica Nucleare\\
  Via Sette Comuni, 56, 10127 Torino TO, Italy}
\affiliation[e]{National Taiwan University\\
  No. 1, Sec. 4, Roosevelt Rd., Taipei 10617, Taiwan (R.O.C.)}
\affiliation[f]{Georgia State University\\
  33 Gilmer Street SE Atlanta, GA 30303, USA}

\emailAdd{yi.chen@cern.ch}
\emailAdd{abaty@rice.edu}
\emailAdd{Dennis.Perepelitsa@colorado.edu}
\emailAdd{chmc7718@colorado.edu}
\emailAdd{jthaler@mit.edu}
\emailAdd{Marcello.Maggi@cern.ch}
\emailAdd{pchang@phys.ntu.edu.tw}
\emailAdd{tasheng@mit.edu}
\emailAdd{ytchien@gsu.edu}
\emailAdd{yenjie@mit.edu}

\abstract{We present the first anti-k$_\mathrm{T}$ jet spectrum and substructure measurements using the archived ALEPH $e^+e^-$ data taken in 1994 at a center of mass energy of $\sqrt{s} = 91.2$~GeV. Jets are reconstructed with the anti-k$_\mathrm{T}$ algorithm with a resolution parameter of 0.4. It is the cleanest test of jets and QCD without the complication of hadronic initial states. The fixed center-of-mass energy also allows the first direct test of pQCD calculation. We present both the inclusive jet energy spectrum and the leading dijet energy spectra, together with a number of substructure observables. They are compared to predictions from PYTHIA6, PYTHIA8, Sherpa, HERWIG, VINCIA, and PYQUEN. None of the models fully reproduce the data. The data are also compared to two perturbative QCD calculations at NLO and with NLL’+R resummation. The results can also serve as reference measurements to compare to results from hadronic colliders. Future directions, including testing jet clustering algorithms designed for future electron-ion collider experiments, will also be discussed.}

\FullConference{%
  41st International Conference on High Energy physics - ICHEP2022\\
  6-13 July, 2022\\
  Bologna, Italy
}

\begin{document}
\maketitle

\section{Introduction}

Experimental signatures similar to those typically attributed to the Quark-gluon plasma have been observed in progressively smaller systems, for example the ridge-like enhancement in two-particle long range correlation in high multiplicity proton-proton collisions at the LHCs~\cite{Khachatryan:2010gv}.  Jet-related observables, however, do not show any signs of modification so far in smaller systems.  It is therefore of interest for jet measurements in the smallest $e^+e^-$ collision systems.
Previous jet measurements in LEP (see e.g. \cite{Buskulic:1995sw}) were limited to earlier generations of jet-finding algorithms and couldn't be directly compared with recent results from hadronic collisions at the LHC and RHIC (e.g. \cite{Abdesselam:2010pt,Salam:2010nqg,Altheimer:2012mn,Larkoski:2017jix}), where the anti-k$_\mathrm{T}$~\cite{Cacciari:2008gp} algorithm is commonly used.
In the current work~\cite{Chen_2022}, we performed the first measurement of energy spectrum and substructure of anti-k$_\mathrm{T}$ jet using the archived data from the ALEPH experiment.

\section{Jet reconstruction and calibration}

Detector-level jets are clustered~\cite{Cacciari:2011ma} from the energy-flow objects with the anti-k$_\mathrm{T}$ algorithm with a distance parameter of 0.4, using the spherical-coordinate variant of the algorithm suitable for the $e^+e^-$ collision system with energy $E$ instead of transverse momentum $k_\mathrm{T}$ and opening angle $\theta$ instead of $\Delta R \equiv \sqrt{\Delta\eta^2+\Delta\phi^2}$.
Due to the small inactive area around the beam pipe, in order to avoid jets that overlap with it, only jets within the acceptance region $0.2\pi < \theta_\text{jet} < 0.8\pi$ are considered.
A hadronic event selection~\cite{Badea:2019vey} is applied to reject events with only electromagnetic interactions.

Jets are calibrated with a multi-step procedure.
As the first step, they are corrected to truth-level jet energies using the archived simulated \textsc{Pythia6}~\cite{Sjostrand:2000wi} sample in different bins of jet direction $\theta_\text{jet}$.
Additional data-simulation difference are derived.  Through leading dijet energies the $e^+$-going and the $e^-$-going sides of the detector are intercalibrated.  The absolute scale is then derived using the event-wide multijet mass, comparing data to simulation, excluding tails which may contain additional effects.  The magnitude of the data-simulation difference goes up to 1\%.

The energy resolution of jets are evaluated first in simulation.  The difference in resolution between data and simulation are measured using the leading dijet energy balance, using the third-leading jet as a handle for systematics on the method.  The measured relative resolution difference is up to 5\% between simulation and data.

For the jet substructure considered in this work, the \textsc{SoftDrop}~\cite{Larkoski:2014wba,Dasgupta:2013ihk} grooming algorithm is used with parameter $(z_\text{cut} = 0.1, \beta = 0.0)$.  The algorithm cleans soft particles scattered at large angles and identifies two subjets, which can be used to define observables.  A few observables are measured: $z_G$, defined as the energy fraction carried by the smaller of the two subjets; $R_G$, the opening angle between the two subjets; $M_G/E$, the invariant mass divided by the two subjets normalized by the energy of the jet; and $M/E$, the invariant mass of the whole jet before the grooming algorithm normalilzed by the energy of the jet.
Results with additional parameter settings and different algorithms are left for future studies.

\section{Analysis}

\begin{figure}[ht!]
    \centering
    \includegraphics[width=0.40\textwidth]{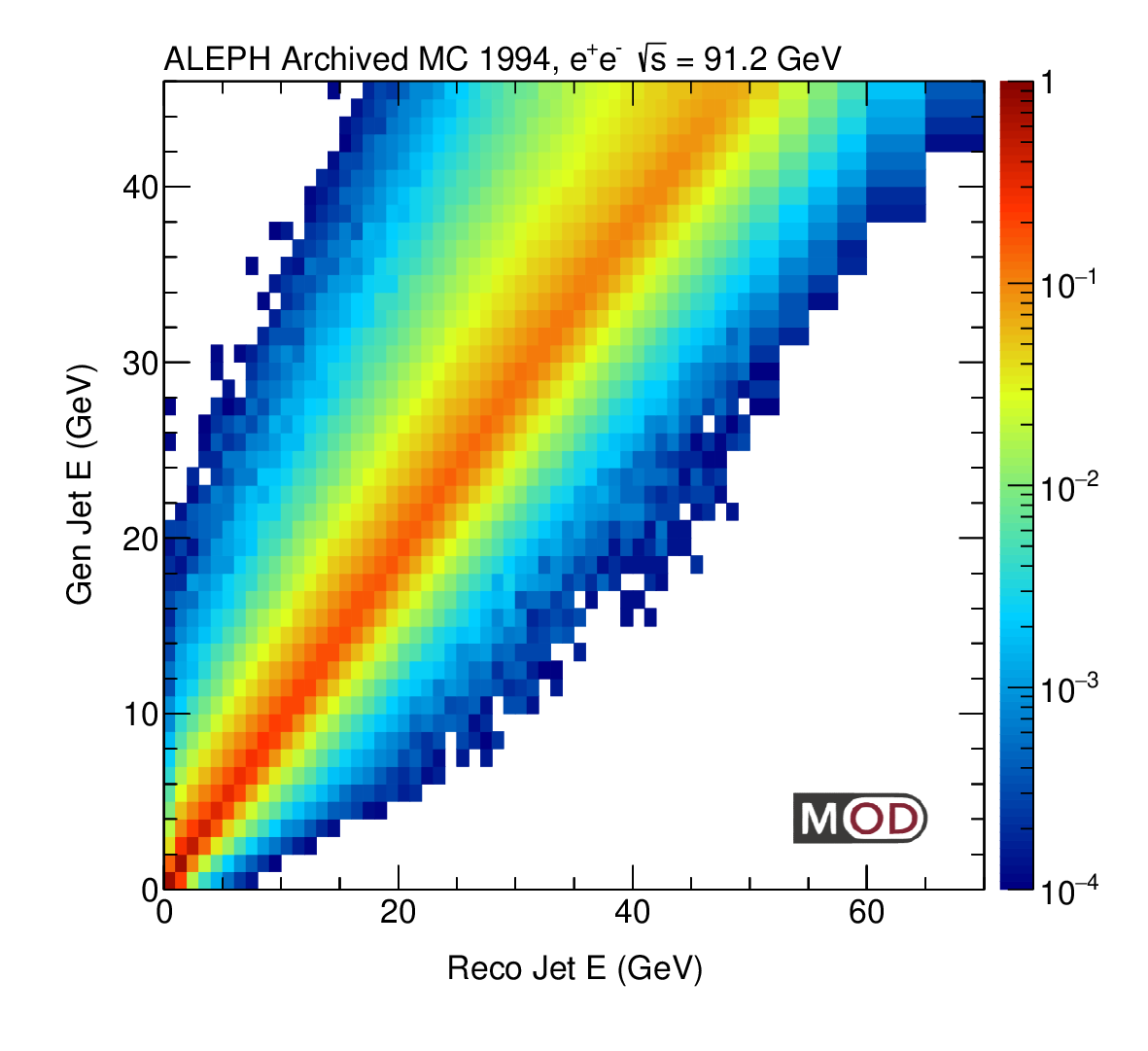}
    \caption{Transfer matrix for inclusive jet energy measurement.}
    \label{Figure:EnergyAnalysis}
\end{figure}

The results are unfolded to account for detector smearing effects.  One example transfer matrix is shown in Fig.~\ref{Figure:EnergyAnalysis} for measurement of jet energy spectrum.  The difference in resolution between data and simulation is incorporated.  Due to the migration in energy between truth-level and detector-level, for substructure measurements the unfolding is done in two dimensions in conjunction with the jet energy.

In order to better quantify the in-cone energy, spectrum of global leading dijet energy is measured.  A correction is derived from simulation to account for cases where some of the energies of the jet is missing due to overlap with the beam direction.

Sources of systematic uncertainties include jet energy scale and resolution, contribution from combinatorial jets not associated with a parton from the hard process, unfolding, and modeling.  For the energy spectra measurements, the dominant source is jet energy scale and resolution, while for the substructure measurements the dominant is from the modeling.

\section{Results}

\begin{figure}[ht!]
    \centering
    \includegraphics[width=0.35\textwidth]{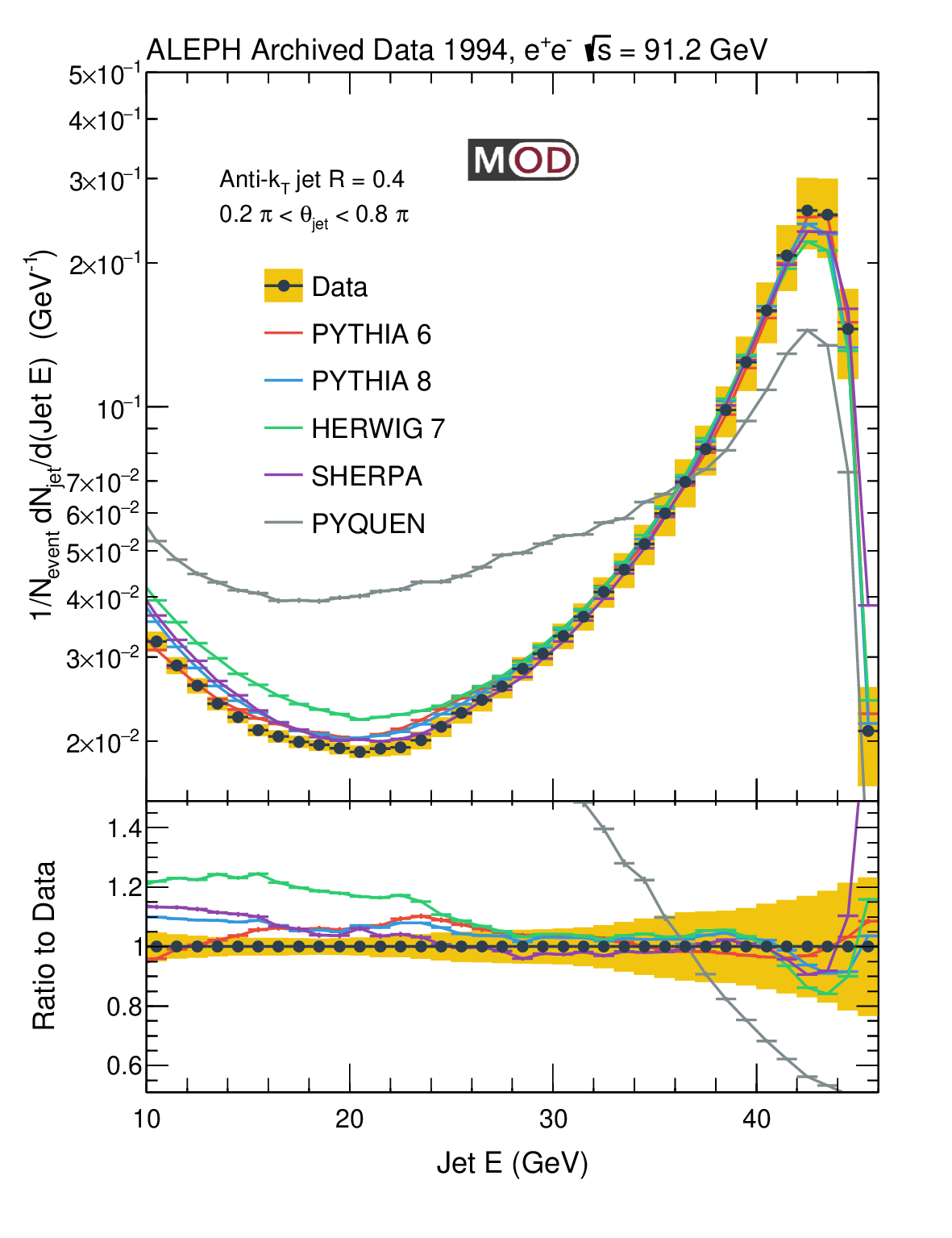}
    \includegraphics[width=0.35\textwidth]{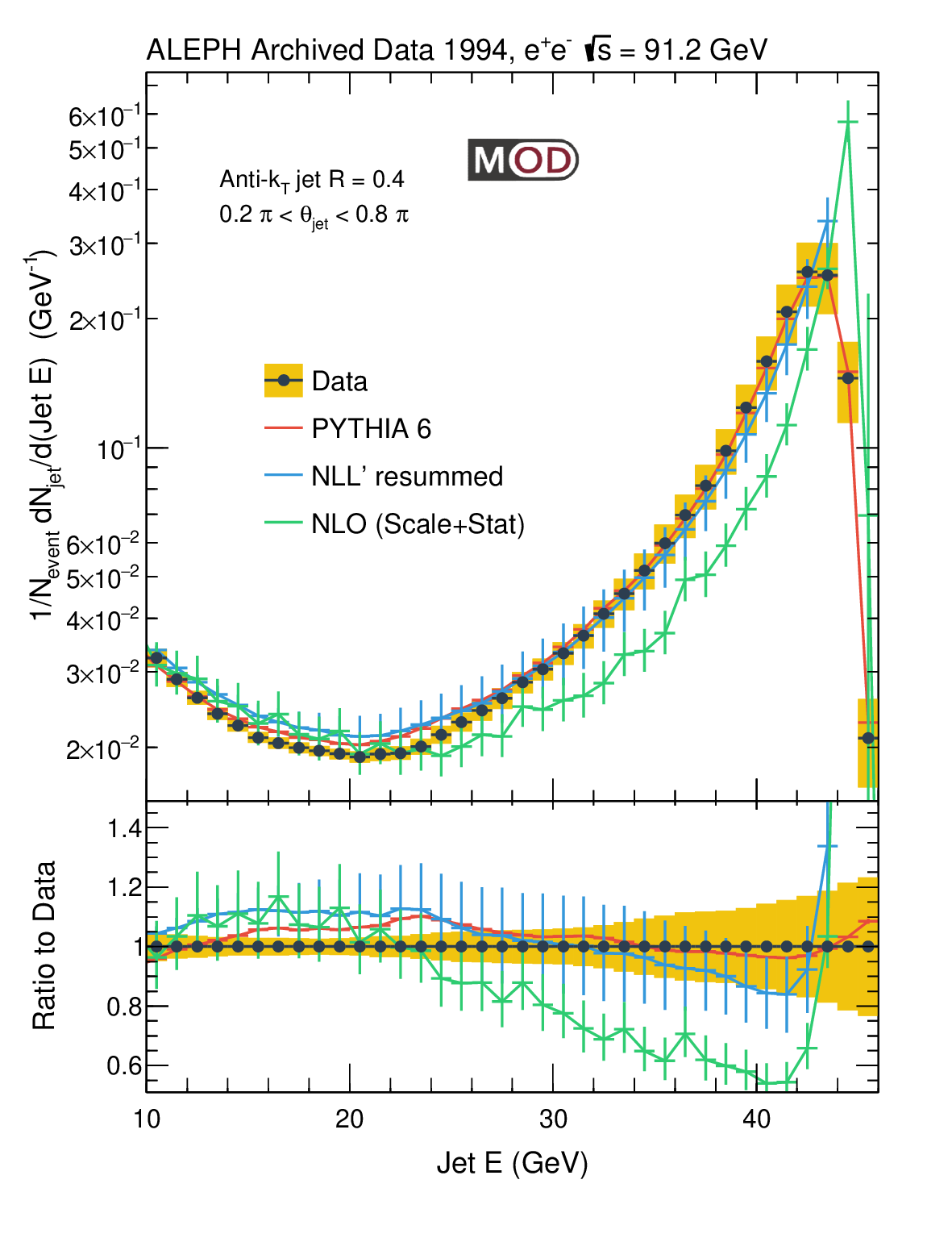}
    \caption{Measured jet energy spectrum compared with Monte-Carlo simulations (left) and NLL'-resummed calculations and parton-level NLO spectrum (right).}
    \label{Figure:ResultEnergy}
\end{figure}

The unfolded energy spectrum is shown in Fig.~\ref{Figure:ResultEnergy}.  In the left panel the measured energy spectrum is compared with \textsc{Pythia}6~\cite{Sjostrand:2000wi}, \textsc{Pythia}8~\cite{Sjostrand:2014zea}, \textsc{Herwig} 7.2.2~\cite{Bellm:2015jjp} and \textsc{Sherpa}~\cite{Gleisberg:2008ta} generators.  The peak above 40 GeV comes from the dominant $e^+e^-\rightarrow q\bar{q}$ process, while the rise at lower energy are from multijet topologies or energies escaped from the leading jets.  The generators generally describe the peak region well, but some discrepancies are observed at low energy.  The data is also compared with NLL'-resummed calculation~\cite{Neill:2021std}, and a general good agreement is observed.  Parton level NLO spectrum is also shown.

\begin{figure}[ht!]
    \centering
    \includegraphics[width=0.35\textwidth]{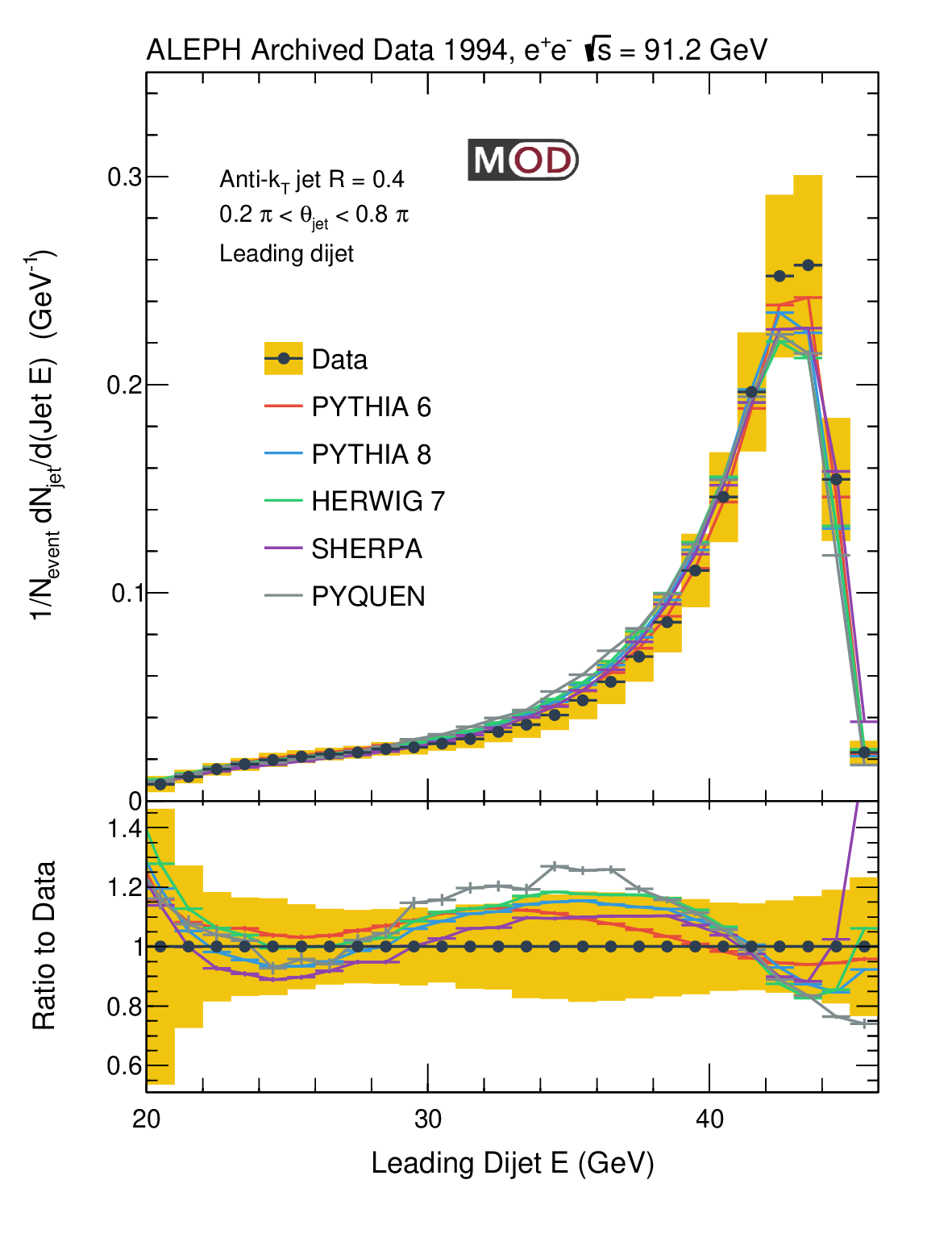}
    \includegraphics[width=0.35\textwidth]{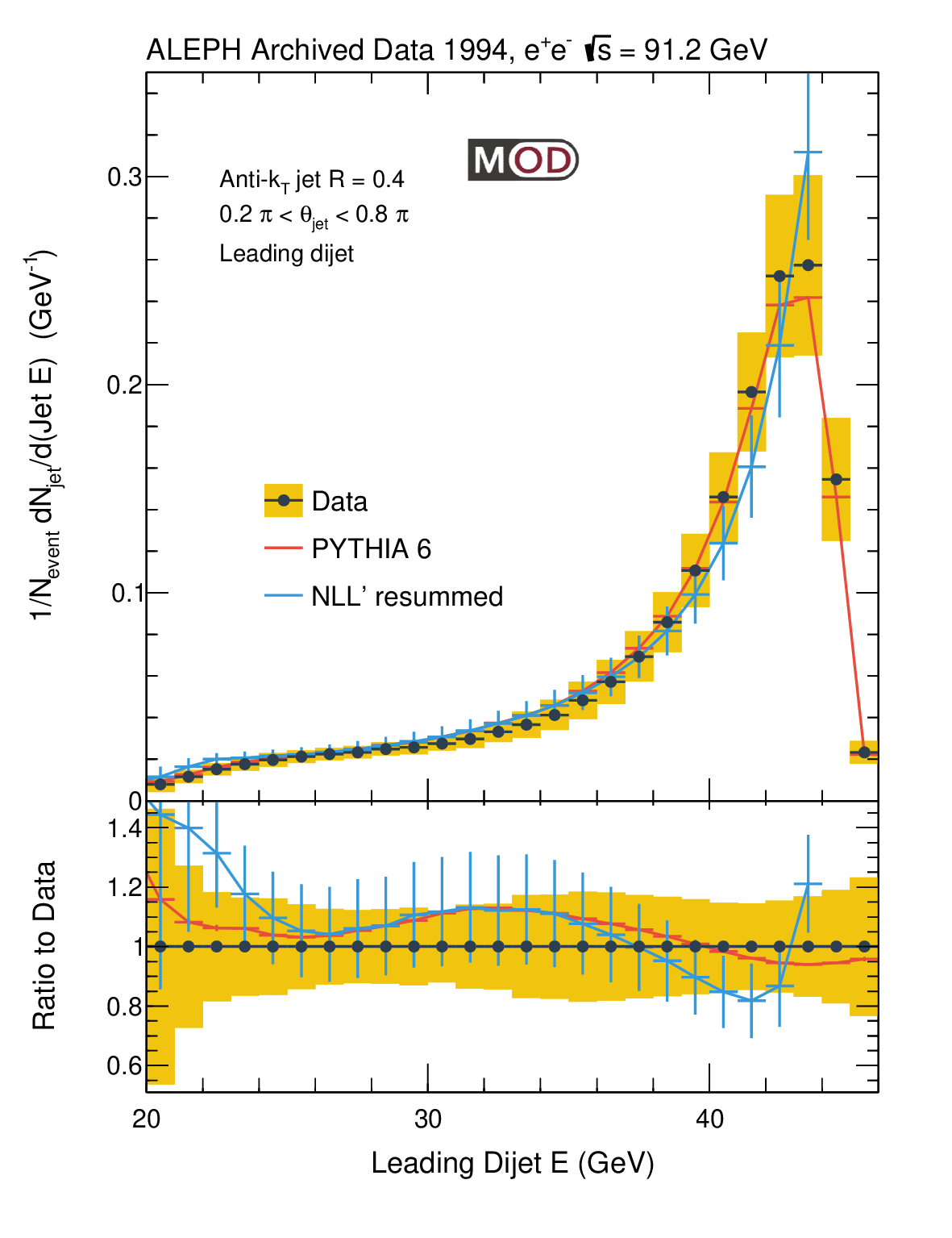}
    \caption{Measured leading dijet jet energy compared with Monte-Carlo simulations (left) and NLL'-resummed calculations (right).}
    \label{Figure:ResultLeadingDijetEnergy}
\end{figure}

In Fig.~\ref{Figure:ResultLeadingDijetEnergy} the measured leading dijet energy is shown.  Even though some trends are seen in the ratio plots, the larger systematic uncertainty prevents a precise statement.  In this case the dominant uncertainty comes from the modeling uncertainty originating from the fact that only one set of detector-level simulation is available to use for studies.

\begin{figure}[ht!]
    \centering
    \includegraphics[width=0.35\textwidth]{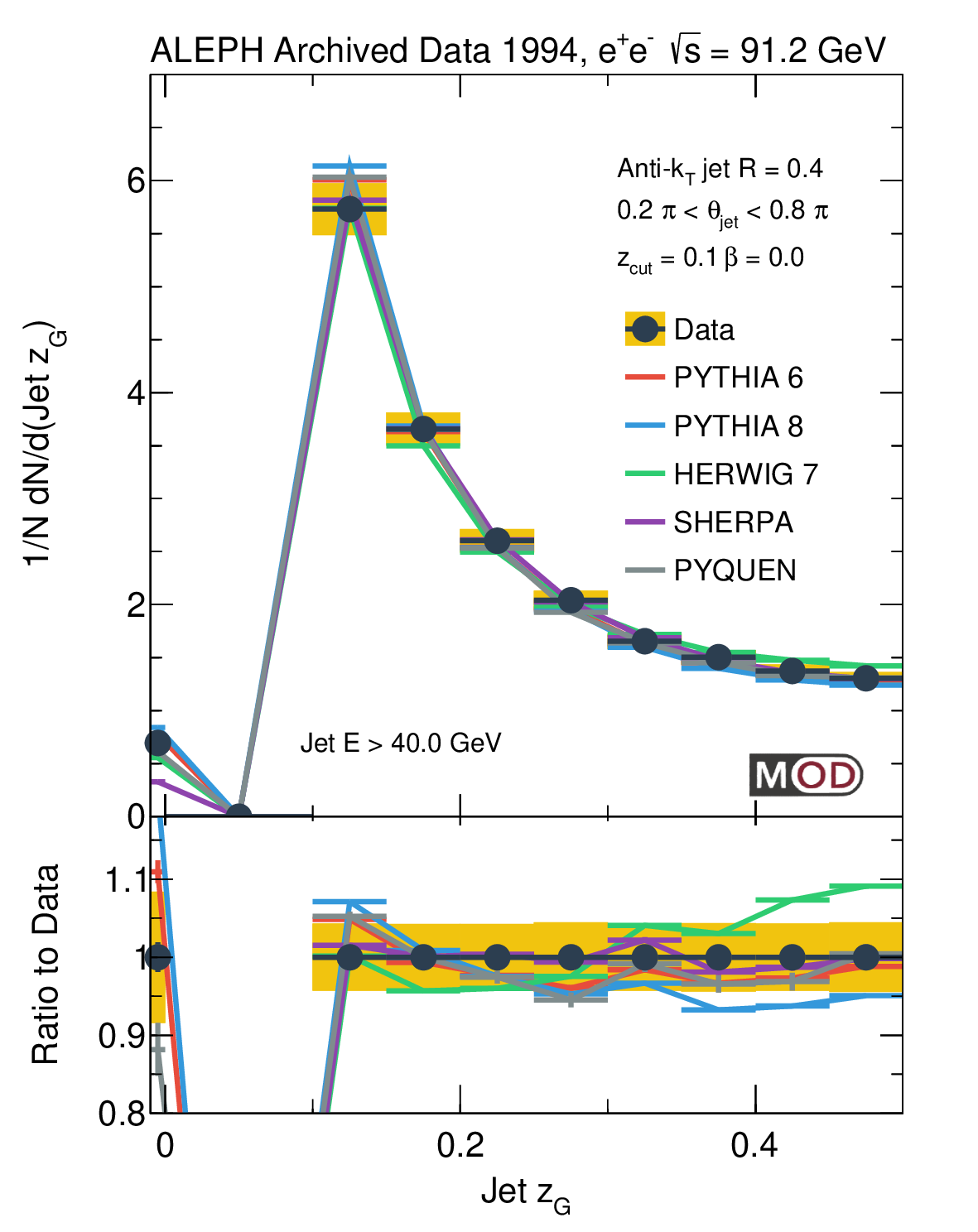}
    \includegraphics[width=0.35\textwidth]{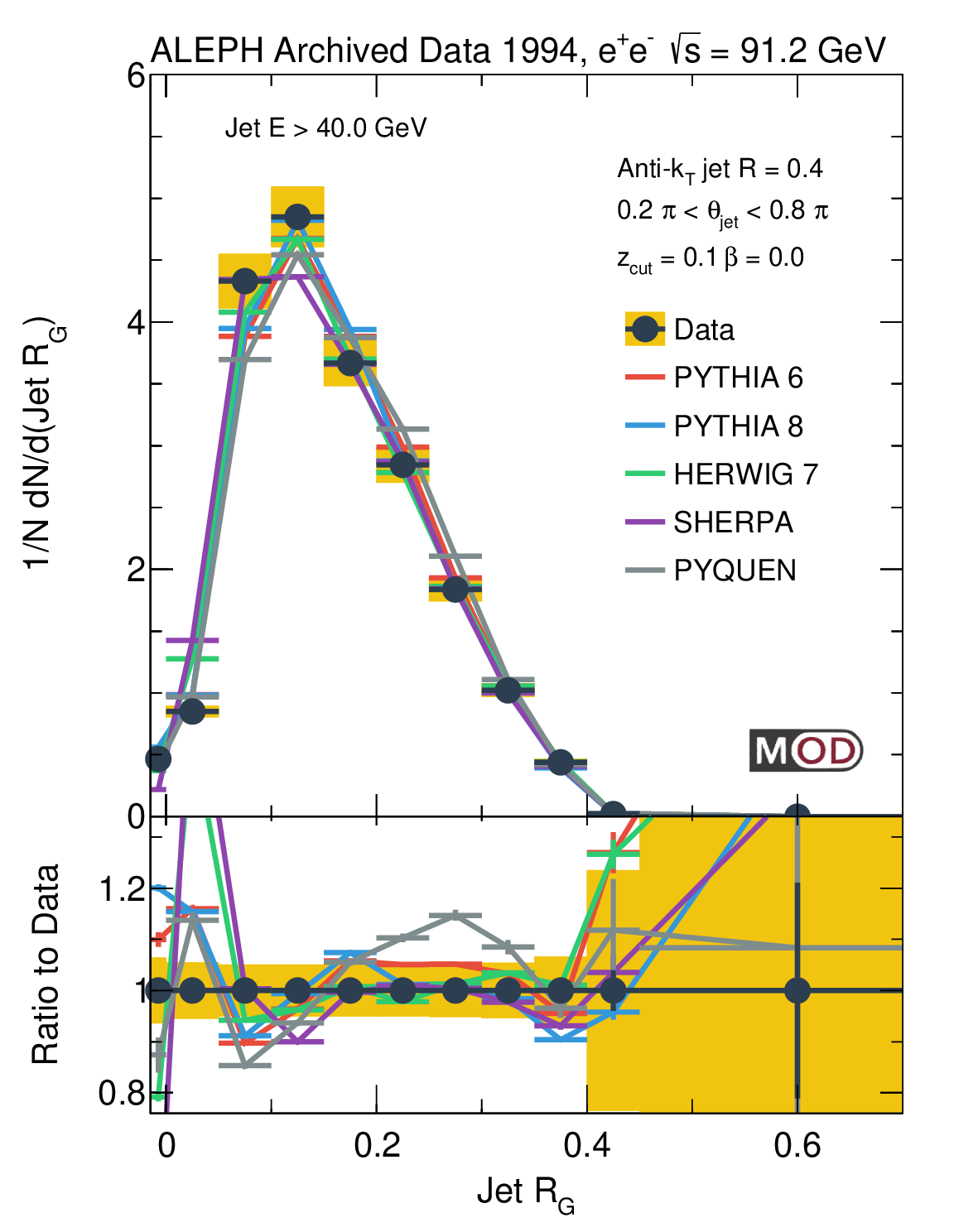}
    \caption{Measured groomed jet energy sharing $z_G$ (left) and $R_G$ (right) for inclusive jets with jet $E > 40$ GeV.  The data is compared with predictions from various generators.}
    \label{Figure:ResultSubstructure}
\end{figure}

The groomed jet energy balance $z_G$ and radius $R_G$ are shown in Fig.~\ref{Figure:ResultSubstructure} for jets with $E > 40$ GeV.  Jets that do not result in two valid subjets through the \textsc{SoftDrop} algorithm are shown in the bin below 0.  The $z_G$ distribution for high energy jets is similar to what has been observed in other collision systems.  There is a deviation up to 10\% between the data and simulations, consistent with the level of discrepancy observed in proton-proton collisions at the LHC.  The bulk of the $R_G$ distribution is well-described by simulations, but there are some discrepancies at the tail of the $R_G$ distribution.

\begin{figure}[ht!]
    \centering
    \includegraphics[width=0.65\textwidth]{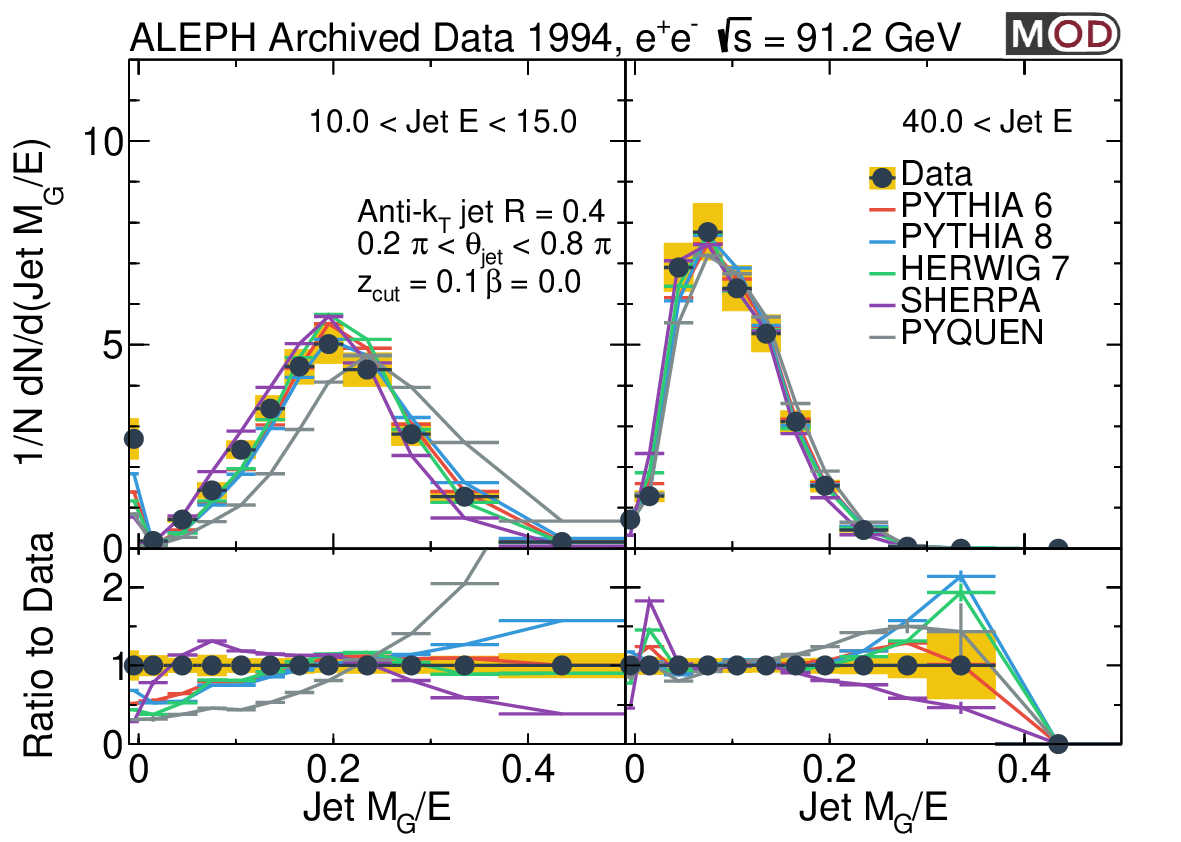}
    \caption{Measured groomed jet mass normalized by jet energy for jets with $10 < E < 15$ GeV (left) and $E > 40$ GeV (right) for inclusive jets.  The data is compared with predictions from various generators.}
    \label{Figure:ResultMass}
\end{figure}

An example of groomed jet mass normalized by jet energy ($M_G/E$) is shown in Fig.~\ref{Figure:ResultMass}.  In the left panel jets with small energy is shown, while in the right panel we show jets with $E > 40$ GeV.  The general behavior is distinct between low and high energies, indicating that the particles are more spread out for low energy jets, whereas the higher energy jets behave more similar to the high $p_T$ jets seen in hadron colliders.  The general shape is nevertheless captured by simulations.  Similar to the observation in $R_G$, the tail of the distributions are less well captured by the simulations.

The full list of results, including results from other jet energy ranges, full jet mass $M/E$ and the sum of leading dijet energy, can be found in ref~\cite{Chen_2022}.

\section{Summary}

We presented the first measurement of the energy and substructure of anti-k$_\mathrm{T}$ jets in $e^+e^-$ collisions using the archived ALEPH data.
The data is generally described by popular event generators, however, some deviations are present for low jet energy and substructure observables.
The result can provide input to QCD theory calculations and event generators.  The calibrated jets also serve as an excellent testing ground for new algorithms in a clean environment without hadronic initial states.

\bibliographystyle{unsrt}
\bibliography{ICHEP2022}

\begin{thebibliography}{10}

\bibitem{Khachatryan:2010gv}
Vardan Khachatryan et~al.
\newblock {Observation of Long-Range Near-Side Angular Correlations in
  Proton-Proton Collisions at the LHC}.
\newblock {\em JHEP}, 09:091, 2010.

\bibitem{Buskulic:1995sw}
D.~Buskulic et~al.
\newblock {Quark and gluon jet properties in symmetric three jet events}.
\newblock {\em Phys. Lett.}, B384:353--364, 1996.

\bibitem{Abdesselam:2010pt}
A.~Abdesselam et~al.
\newblock {Boosted Objects: A Probe of Beyond the Standard Model Physics}.
\newblock {\em Eur. Phys. J. C}, 71:1661, 2011.

\bibitem{Salam:2010nqg}
Gavin~P. Salam.
\newblock {Towards Jetography}.
\newblock {\em Eur. Phys. J. C}, 67:637--686, 2010.

\bibitem{Altheimer:2012mn}
A.~Altheimer et~al.
\newblock {Jet Substructure at the Tevatron and LHC: New results, new tools,
  new benchmarks}.
\newblock {\em J. Phys. G}, 39:063001, 2012.

\bibitem{Larkoski:2017jix}
Andrew~J. Larkoski, Ian Moult, and Benjamin Nachman.
\newblock {Jet Substructure at the Large Hadron Collider: A Review of Recent
  Advances in Theory and Machine Learning}.
\newblock {\em Phys. Rept.}, 841:1--63, 2020.

\bibitem{Cacciari:2008gp}
Matteo Cacciari, Gavin~P. Salam, and Gregory Soyez.
\newblock {The anti-$k_t$ jet clustering algorithm}.
\newblock {\em JHEP}, 04:063, 2008.

\bibitem{Chen_2022}
Yi~Chen et~al.
\newblock Jet energy spectrum and substructure in $e^+e^-$ collisions at 91.2
  {GeV} with {ALEPH} archived data.
\newblock {\em JHEP}, 2022(6):008, 2022.

\bibitem{Cacciari:2011ma}
Matteo Cacciari, Gavin~P. Salam, and Gregory Soyez.
\newblock {FastJet User Manual}.
\newblock {\em Eur. Phys. J. C}, 72:1896, 2012.

\bibitem{Badea:2019vey}
Anthony Badea, Austin Baty, Paoti Chang, Gian~Michele Innocenti, Marcello
  Maggi, Christopher Mcginn, Michael Peters, Tzu-An Sheng, Jesse Thaler, and
  Yen-Jie Lee.
\newblock {Measurements of two-particle correlations in $e^+e^-$ collisions at
  91 GeV with ALEPH archived data}.
\newblock {\em Phys. Rev. Lett.}, 123(21):212002, 2019.

\bibitem{Sjostrand:2000wi}
Torbj{\"o}rn Sj{\"o}strand, Patrik Eden, Christer Friberg, Leif Lonnblad,
  Gabriela Miu, Stephen Mrenna, and Emanuel Norrbin.
\newblock {High-energy physics event generation with PYTHIA 6.1}.
\newblock {\em Comput. Phys. Commun.}, 135:238--259, 2001.

\bibitem{Larkoski:2014wba}
Andrew~J. Larkoski, Simone Marzani, Gregory Soyez, and Jesse Thaler.
\newblock {Soft Drop}.
\newblock {\em JHEP}, 05:146, 2014.

\bibitem{Dasgupta:2013ihk}
Mrinal Dasgupta, Alessandro Fregoso, Simone Marzani, and Gavin~P. Salam.
\newblock {Towards an understanding of jet substructure}.
\newblock {\em JHEP}, 09:029, 2013.

\bibitem{Sjostrand:2014zea}
Torbj\"orn Sj\"ostrand, Stefan Ask, Jesper~R. Christiansen, Richard Corke,
  Nishita Desai, Philip Ilten, Stephen Mrenna, Stefan Prestel, Christine~O.
  Rasmussen, and Peter~Z. Skands.
\newblock {An introduction to PYTHIA 8.2}.
\newblock {\em Comput. Phys. Commun.}, 191:159--177, 2015.

\bibitem{Bellm:2015jjp}
Johannes Bellm et~al.
\newblock {Herwig 7.0/Herwig++ 3.0 release note}.
\newblock {\em Eur. Phys. J. C}, 76(4):196, 2016.

\bibitem{Gleisberg:2008ta}
T.~Gleisberg, Stefan. Hoeche, F.~Krauss, M.~Schonherr, S.~Schumann, F.~Siegert,
  and J.~Winter.
\newblock {Event generation with SHERPA 1.1}.
\newblock {\em JHEP}, 02:007, 2009.

\bibitem{Neill:2021std}
Duff Neill, Felix Ringer, and Nobuo Sato.
\newblock {Leading jets and energy loss}.
\newblock {\em JHEP}, 07:041, 2021.

\end{thebibliography}

\end{document}